\documentclass[twocolumn,showpacs,amsmath,amssymb]{revtex4}
\usepackage{graphicx}
\newcommand{\ltsim}{\protect\raisebox{-0.5ex}{$\:\stackrel{\textstyle <}{\sim}\:$}}

\begin{document}

\title{Entropy of Factorized Snapshot Data for Two-Dimensional Classical Spin Models}
\author{Hiroaki Matsueda and Dai Ozaki}
\affiliation{Sendai National College of Technology, Sendai 989-3128, Japan}
\date{\today}
\begin{abstract}
We reexamine the snapshot entropy of the Ising and three-states Potts models on the $L\times L$ square lattice. Focusing on the factorization of the snapshot matrix, we find that the entropy at $T_{c}$ scales asymptotically as $S\sim(c/3)\ln L$ consistent with the entanglement entropy in one-dimensional quantum critical systems. This nontrivial consistency strongly supports that the snapshot entropy after the factorization really represents the holographic entanglement entropy. On the other hand, the anomalous scaling $S_{\chi}\sim\chi^{\eta}\ln\chi$ for the coarse-grained snpshot entropy is retained even after the factorization. These fearures are considered to originate from the fact that the largest singular value of the snapshot matrix is regulated by the factorization.
\end{abstract}
\maketitle

\section{Introduction}

Quantum entanglement and holography are two fundamental concepts in current theoretical physics. The entropy of the entanglement has information similar to the two-point correlation function of scaling operators, and is recognized to be a new order parameter efficient for detecting quantum-ordered states. On the other side, the holography provides us a method of how to transform the quantum entanglement into classical quantities. A representative holographic connection is the so-called Ryu-Takayanagi formula in the string theory that transforms the entanglement entropy into the minimal surface area on the classical side with negative curvature~\cite{Ryu}. The negative curvature space is characterized by the presence of various length scales, and in other words the holography seems to be a method of how to efficiently embed the quantum data into the classical memory space.

Although the holography has been mainly examined in string-theory community, one of the authors (HM) has also proposed a possible alternative holographic theory appearing in the statistical physics~\cite{Matsueda}. Interestingly, the theory is based on the image processing of spin snapshot data in the classical two-dimensional (2D) Ising and three-states Potts models. The image processing is done by the singular value decomposition (SVD), and because of the universal nature of the singular values, the information entropy defined from the singular values shows a universal feature that reminds us with the entanglement entropy in 1D quantum critical systems. We call this entropy as snapshot entropy. In the previous paper, this correspondence was considered to originate from the Suzuki-Trotter decomposition of the corresponding 1D transverse-field quantum Ising model~\cite{Suzuki}. SVD can also be applied to both of classical and quantum systems, and the function of SVD as length-scale decomposition works in both systems. Thus, we think that this common feature can detects the quantum-classical correspondence.

However, there are still many questions for the physical meaning of the snapshot entropy, although related works appear recently~\cite{Imura,Matsueda2,Matsueda3}. One of mysteries is why the snapshot entropy at $T_{c}$ behaves as $S\sim\ln L$. According to the conformal field theory (CFT), the entanglement entropy in 1D critical systems is obtained as
\begin{eqnarray}
S=\frac{c}{3}\ln L+c_{1}, \label{CFT}
\end{eqnarray}
where $c$ is the central charge of corresponding CFT and $c_{1}$ is a non-universal positive constant~\cite{Holzhey,Calabrese}. If the snapshot entropy is a key quantity of holography, we expect that the amount of information should conserve after the quantum-classical correspondence, and thus the coefficient $c/3$ should appear.

A hint to resolve this problem appears in the author's recent work on the snapshot entropy of fractal images~\cite{Matsueda2}. Fractal images are ideal Ising snapshots at $T_{c}$ owing to their self-similarities. Here, a fractal is defined by the tensor product of the $h\times h$ unit cell matrix $H$. Then, the tensor product of $N$ copies of $H$, $M=H\otimes H\otimes\cdots\otimes H\otimes H$, represents the fractal with $N$ different length scales. An important point is that the unit cell matrix $H$ should be 'factorized' so that the tensor product properly creates the desired fractal image. Then, the entropy is given by $S\propto N\propto \ln L$ for the total system size $L=h^{N}$, and we know that the snapshot entropy counts the number of the different cluster sizes. Therefore, the factorization is quite important for the scaling formula. For example, let us consider the unit cell matrix of the Sierpinski carpet given by
\begin{eqnarray}
H=\left(\begin{array}{ccc}1&1&1\\ 1&0&1\\ 1&1&1\end{array}\right). \label{unit}
\end{eqnarray}
The tensor product of two copies of $H$ is given by
\begin{eqnarray}
H\otimes H=\left(\begin{array}{ccc|ccc|ccc}
1&1&1&1&1&1&1&1&1 \\
1&0&1&1&0&1&1&0&1 \\
1&1&1&1&1&1&1&1&1 \\ \hline
1&1&1&0&0&0&1&1&1 \\
1&0&1&0&0&0&1&0&1 \\
1&1&1&0&0&0&1&1&1 \\ \hline
1&1&1&1&1&1&1&1&1 \\
1&0&1&1&0&1&1&0&1 \\
1&1&1&1&1&1&1&1&1
\end{array}\right),
\end{eqnarray}
and clearly the level-$2$ fractal structure emerges. However, we cannot exchange $0$ and $1$, since the matrix elements $M$ are all zero except for the central pixel. In this viewpoint, the previous works did not take care about the factorization. Therein, the up and down spins of the Ising model were taken to be $+1$ and $-1$. We need to reconsider whether this selection really produces the factorized form or not. The purpose of this paper is to confirm the role of the factorization on the snapshot entropy near $T_{c}$. Then, the factorization will make it clear to know the close connection among the central charge, hierarchical cluster spins, and the entropy formula.

We will actually find that the factorization is crucial to the proper scaling relation of the snapshot entropy at $T_{c}$. The proper selection of the matrix elements would be $0$ and $1$ for up and down spins of the Ising model. The selection for the Potts model would be $0$, $1$, and $2$. Then, the entanglement entropy formula given by Eq.~(\ref{CFT}) is actually reproduced up to the coefficient $c/3$ by the calculation of the snapshot entropy. We will also mention the largest eigenvalue spectrum, and this also shows similar scaling behavior. We will point out that the largest singular value should be properly regulated by the factorization.

The organization of this paper is as follows. In the next section, we discuss the role of factorization on the SVD spectrum. In Sec.~III, the numerical data and the scaling analysis for the 2D Ising model are presented, and we will give a theoretical interpretation of the scaling based on CFT and the Suzuki-Trotter decompostion. In Sec.~IV, we examine close connection between the tensor product of the factorized snapshot matrix and the cluster distribution. We also present the finite-$\chi$ scaling in Sec.~V. The reliability of our scaling is examined in Sec.~VI by taking the three-states Potts model with the different central charge. Sec.~VII is devoted to the summary part.

\section{Factorization of Snapshot Data}

\subsection{Role of the largest singular value on the snapshot entropy}

The Hamiltonian of the Ising model is defined by
\begin{eqnarray}
H=-J\sum_{\left<i,j\right>}\sigma_{i}\sigma_{j}.
\end{eqnarray}
where $\sigma_{i}=\pm 1$ and the sum runs over the nearest neighbor lattice sites $\left<i,j\right>$, and $J (>0)$ is the exchange interaction. We consider the square lattice under the periodic boundary condition, and the system size is taken to be $L\times L$. The critical temperature is known to be $T_{c}/J=2/\ln (1+\sqrt{2})=2.2692$. The central charge of CFT is $c=1/2$.

We regard a snapshot (a spin configuration) as the following matrix
\begin{eqnarray}
M_{n}(x,y)=\sigma_{i}, \label{non}
\end{eqnarray}
with $i=(x,y)$. The index $n$ means 'non'-factorization. To obtain the snapshot, we perform Monte Carlo (MC) simulation by the Metropolis algorithm. Starting with the temperature $T=3.0J$ ($T=1.5J$), we gradually reduce (increase) $T$ by $\Delta T=0.01J$, and take $10^{4}\sim 10^{5}$ MC steps for convergence at each $T$. Here, one MC step counts $L\times L$ local updates. At each $T$, we calculate the snapshot entropy with the help of SVD. We also take the Swendsen-Wang algorithm for more sophisticated scaling analysis (Figs.~\ref{fig2}, \ref{fig3}, \ref{fig4}, and~\ref{fig5}). Then, we take $\Delta T=0.001\times T_{c}$, and $10^{4}$ MC steps. Here, one MC step corresponds to all possible cluster flips for a given spin configuration.

The factorization of Eq.~(\ref{non}) is defined by
\begin{eqnarray}
M_{f}(x,y)=\frac{1}{2}\left(\sigma_{i}+1\right). \label{shift}
\end{eqnarray}
In the ferromagnetic phase, we need to take care about many zeros of matrix elements, since in such a case, SVD and entropy calculation become unstable. If the total magnetization $\sum_{i}\sigma_{i}$ is positive, we replace $M(x,y)=-1$ into $0$. If the magnetization is nagative, we take $M(x,y)=1\rightarrow 0$ and $M(x,y)=-1\rightarrow 1$. We are particularly interested in the case of Eq.~(\ref{shift}).

As already mentioned, the snapshot data $M(x,y)$ can be regarded as a matrix. To extract the universal information of the matrix data, we apply SVD to $M(x,y)$. The SVD is defined by
\begin{eqnarray}
M(x,y)=\sum_{n=1}^{L}U_{n}(x)\sqrt{\Lambda_{n}}V_{n}(y). \label{SVD}
\end{eqnarray}
where $\sqrt{\Lambda_{n}}$ is the $n$th singular value, and $U_{n}(x)$ and $V_{n}(y)$ are $n$th column unitary matrices. The square of the sigular value, $\Lambda_{n}$, is equal to the eigenvalue of the density matrix defined by
\begin{eqnarray}
\rho=MM^{\dagger}.
\end{eqnarray}
We align the eigenvalues so that $\Lambda_{1}\ge\Lambda_{2}\ge\cdots\ge\Lambda_{L}$. Each eigenvalue $\Lambda_{n}$ is normalized to be
\begin{eqnarray}
\lambda_{n}=\frac{\Lambda_{n}}{\sum_{n=1}^{L}\Lambda_{n}}.
\end{eqnarray}
Then, the snapshot entropy is defined by
\begin{eqnarray}
S=-\sum_{n=1}^{L}\lambda_{n}\ln\lambda_{n}. \label{entropy}
\end{eqnarray}
In the following, we examine the $T$ and $L$ dependences of this entropy $S$.

Before going into details of numerical data, we mention the density matrices for both factorized and non-factorized snapshot data, $M_{f}$ and $M_{n}$, respectively. These data are related with
\begin{eqnarray}
M_{f}=\frac{1}{2}\left(M_{n}+B\right),
\end{eqnarray}
where the $L\times L$ background matrix $B$ is defined by
\begin{eqnarray}
B=\left(\begin{array}{ccc}
1&\cdots&1 \\
\vdots&\ddots&\vdots \\
1&\cdots&1
\end{array}\right). \label{b}
\end{eqnarray}
Then the density matrix for $M_{f}$ is given by
\begin{eqnarray}
\rho_{f}=\frac{1}{4}\rho_{n} + \frac{1}{4}\left(BM_{n}^{\dagger}+M_{n}B\right) + \frac{1}{4}LB, \label{rhof}
\end{eqnarray}
where we have used $B^{2}=LB$. The second term, $\left(BM_{n}^{\dagger}+M_{n}B\right)/4$, is very small for paramagnetic states and near $T_{c}$, since $M_{n}$ contains both of $\pm 1$ components and the matrix product almost cancels out. Therefore, the eigenvalues of $\rho_{f}$ are quite different from those of $\rho_{n}$ as long as the background matrix $B$ is not the unit matrix. This means that the factorization is characterized by $B$ and in the present case, $B$ in Eq.~(\ref{b}) affects the final result very much.

According to the recent paper~\cite{Imura}, the non-factorized density matrix $\rho_{n}$ corresponds to the spin-spin correlator
\begin{eqnarray}
\rho_{n}(x,y)\sim G(x-y)\propto\left|x-y\right|^{-\eta},
\end{eqnarray}
with the anomalous dimension $\eta=1/4$, and is diagonalized by the Fourier transformation. The background matrix $B$ is also diagonalized by the Fourier transformation, and the third term in Eq.~(\ref{rhof}) contributes to the uniform ($k=0$) component. Thus if the second term in Eq.~(\ref{rhof}), $(BM_{n}^{\dagger}+M_{n}B)/4$, does not affect so much for the final result, the factorization makes the maximum eigenvalue increased. After the normalization of a set of all the eigenvalues, the snapshot entropy decreases, since the probability of only the largest singular-value state increases. As we have already mentioned in the introduction, the entropy without the factorization is given by $S\sim\ln L$. Thus, it is natural that a factor less than unity appears in front of the logarithmic term.

The above statement only guarantees the presence of the small factor, and we do not mention anymore about its physical origin. However, it is found in the entropy calculation by SVD that the mangitude of the largest eivenvalue $\lambda_{1}$ is crucial for determining the total entropy value including the central charge. This point has been already clarified on the quantum side~\cite{Lefevre}. According to CFT, the logarithm of the largest eigenvalue is a half of the total entanglement entropy
\begin{eqnarray}
-\ln\lambda_{1}=\frac{S}{2}=\frac{c}{6}\ln L, \label{spectrum}
\end{eqnarray}
and the central charge appears through the scaling formula of $S$. We hypothesize that this property is also reflected to the classical side. If $\lambda_{1}$ is the uniform component of the Fourier-transformed correlator $G(k=0)$, $-\ln\lambda_{1}$ is almost equal to the definition of the entanglement entropy in terms of Calabrese-Cardy's approach~\cite{Calabrese}. Maybe the factorization facilitates to pick up the uniform component, although in the present state a role of additional constant factor by the factorization on $\lambda_{1}$ is still missing. We will examine whether this scaling equation is also satisfied on the classical side as well as the entanglement entropy formula.

\subsection{Representation of snapshot matrix and non-zero singular values}

Let us further refer to the representation dependence on the SVD spectra. We focus on whether a particular choice of representation can detect fine structures of the spin congifuration. Then, we would like to confirm that the background matrix is necessary for the representation. If this is correct, the background matrix leads to the important result that the choice is related to a proper regulation of the largest eigenvalue. For these purposes, we consider the following examples ($\alpha\ne 1$)
\begin{eqnarray}
M_{1}(\alpha)=\left(\begin{array}{cccc}1&1&1&1\\ \alpha&\alpha&\alpha&\alpha\end{array}\right) , \; 
M_{2}(\alpha)=\left(\begin{array}{cccc}1&1&\alpha&\alpha\\ \alpha&\alpha&1&1\end{array}\right).
\end{eqnarray}
Here, we regard an alignment of $1$ (or $\alpha$) as a spin string, and in the case of $M_{2}$ the strings of $1$ and $\alpha$ are classically entangled. At first let us take $\alpha=0$. Then, their partial density matrices are given by
\begin{eqnarray}
M_{1}(0)M_{1}(0)^{\dagger}=\left(\begin{array}{cc}4&0\\ 0&0\end{array}\right) , \; 
M_{2}(0)M_{2}(0)^{\dagger}=\left(\begin{array}{cc}2&0\\ 0&2\end{array}\right).
\end{eqnarray}
Thus, the matrix $M_{2}(0)$ has larger entropy than that for the matrix $M_{1}(0)$. This means that the complexity of string configuration affects the entropy value. However, if we replace $0$ to $-1$, the situation changes. Actually,
\begin{eqnarray}
M_{1}(-1)M_{1}(-1)^{\dagger}=M_{2}(-1)M_{2}(-1)^{\dagger}=\left(\begin{array}{cc}4&-4\\ -4&4\end{array}\right),
\end{eqnarray}
and then we can not identify the essential difference between $M_{1}(-1)$ and $M_{2}(-1)$ by the snapshot entropy that is only dependent to the eigenvalues of this matrix. We think that the excess entropy induced by a larger-entropy representation is an artifact, and we should determine the minimum-entropy state. In this strategy, negative components are not willing, since the effect of the negative sign on the partial density matrix sometimes vanishes.

The uniform shift of the matrix elements, like the backgroud matrix $B$ in the previous subsection, is an efficient way of taking account of the above strategy. We introduce the shifted matrix for $M_{1}$ as
\begin{eqnarray}
M_{1}(\alpha;\Delta)&=&M_{1}(\alpha)-\Delta B \nonumber \\
&=&\left(\begin{array}{cccc}1-\Delta&1-\Delta&1-\Delta&1-\Delta\\ \alpha-\Delta&\alpha-\Delta&\alpha-\Delta&\alpha-\Delta\end{array}\right).
\end{eqnarray}
Then, the partial density matrix is given by
\begin{eqnarray}
&& M_{1}(\alpha;\Delta)M_{1}(\alpha;\Delta)^{\dagger} \nonumber \\
&&\;\; =\left(\begin{array}{cc}4(1-\Delta)^{2}&4(1-\Delta)(\alpha-\Delta)\\ 4(1-\Delta)(\alpha-\Delta)&4(\alpha-\Delta)^{2}\end{array}\right),
\end{eqnarray}
and we should take $\Delta=\alpha$ for the minimum entropy state. Finally, we obtain
\begin{eqnarray}
M_{1}(\alpha;\alpha)=(1-\alpha)M_{1}(0),
\end{eqnarray}
and taking $\alpha=0$ is the simplest selection of the matrix elements. This is one interpretation of taking $0$ matrix elements.

\section{Numerical Results for Snapshot Entropy Before and After Factorization}

\subsection{Single Snapshot Entropy}

\begin{figure}[htbp]
\begin{center}
\includegraphics[width=7cm]{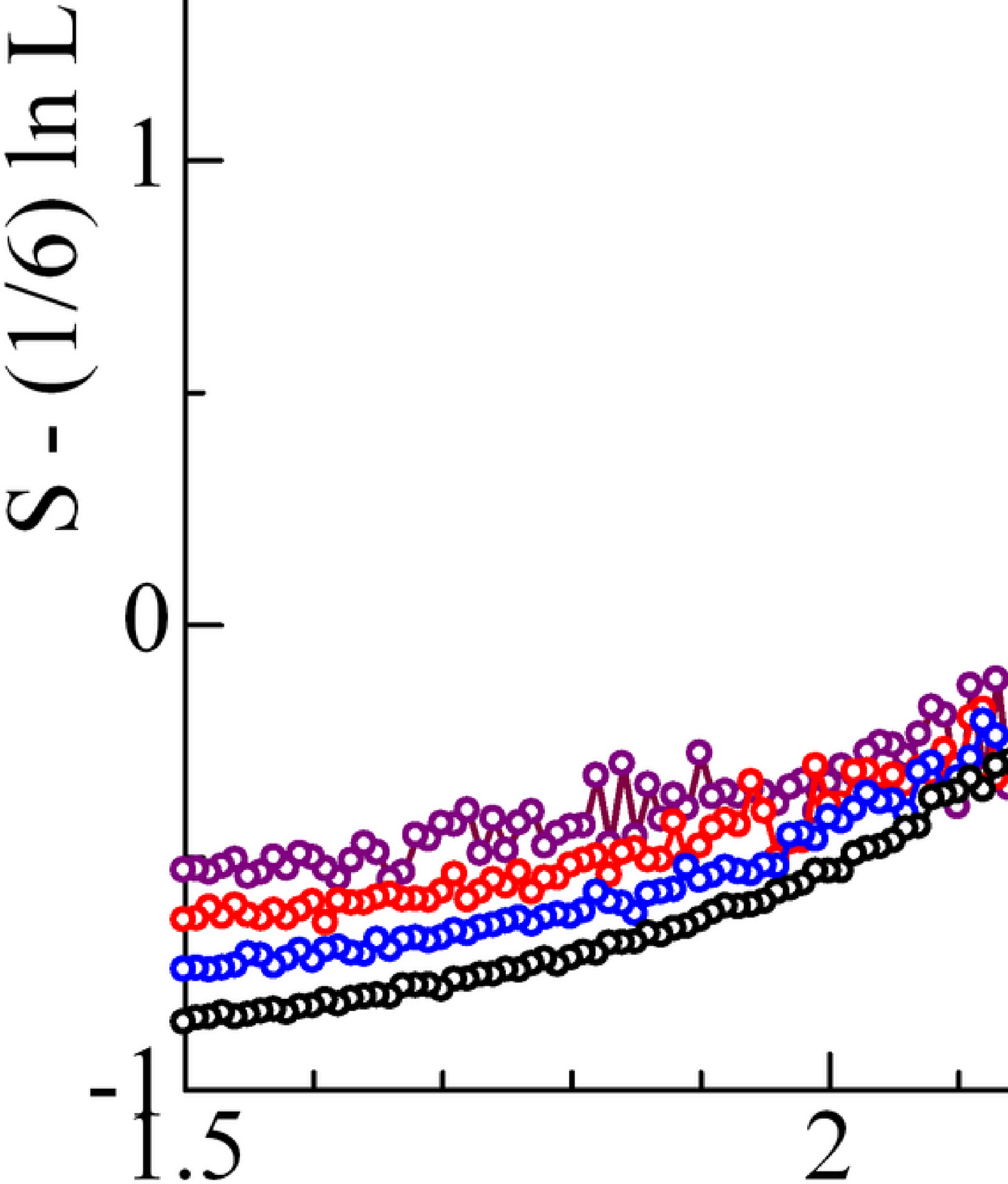}
\includegraphics[width=7cm]{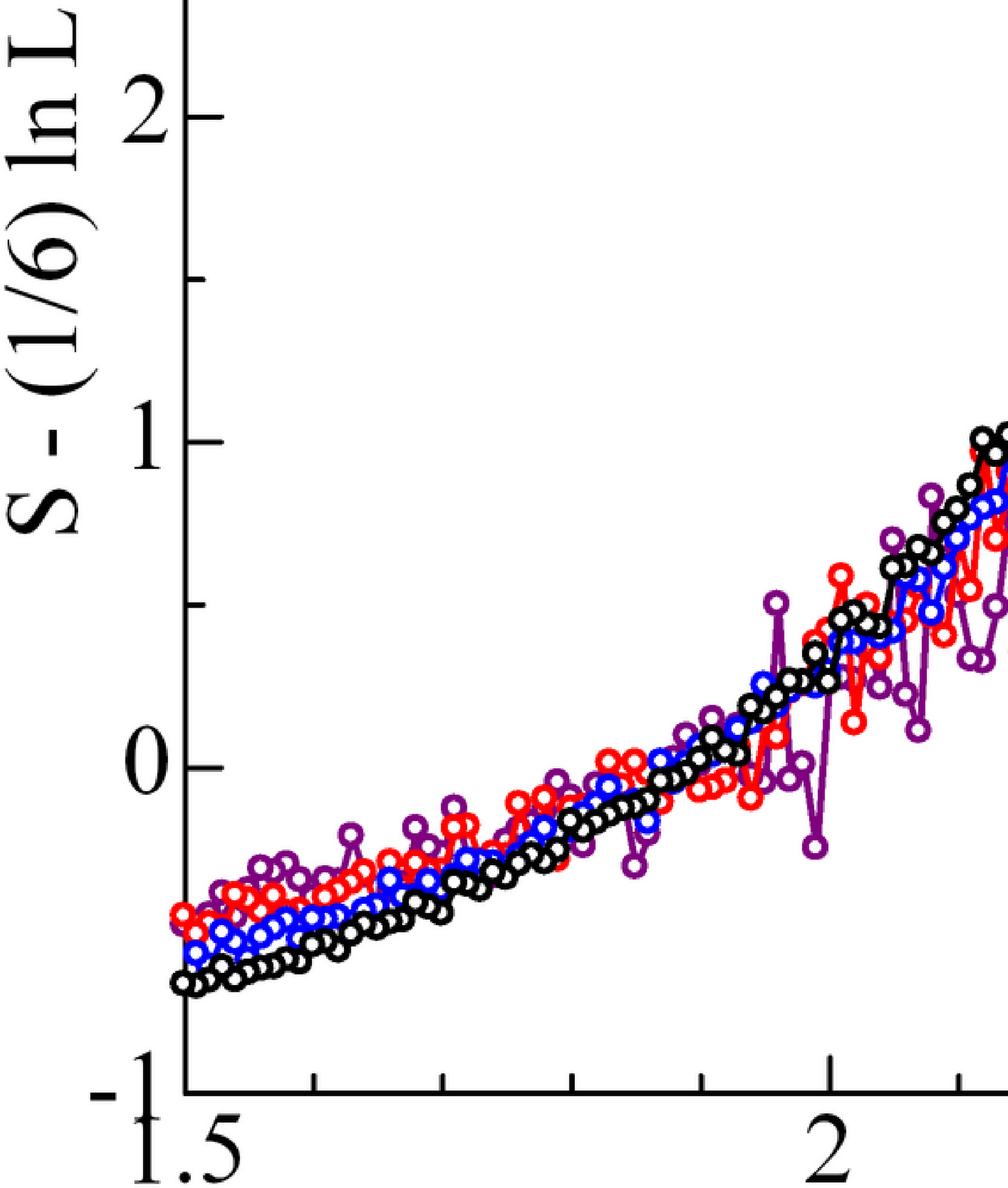}
\end{center}
\caption{(Color online) Temperature and system-size dependence on $S-(c/3)\ln L$: $L=32=2^{5}$ (purple), $L=64=2^{6}$ (red), $L=128=2^{7}$ (blue), and $L=256=2^{8}$ (black). A dashed vertical line is a guide to $T_{c}$. The upper panel (a) is numerical result for $M(x,y)=\left(\sigma_{i}+1\right)/2$ with $i=(x,y)$, and the lower panel (b) is for $M(x,y)=\sigma_{i}$.}
\label{fig1}
\end{figure}

Let us examine the numerical results for the Ising model. Figure~\ref{fig1}(a) shows $S-(c/3)\ln L$ as functions of $T$ and $L$. We have used single snapshot for each $T$, and does not take statistical avarage. The statistical error becomes smaller with increasing $L$ owing to the self-avaraging feature of the snapshot entropy, but at the same time we would like to find the temperature range in which critical fluctuation is mostly enhanced. We will later also present the sample avarage. In Fig.~\ref{fig1}(a), we have used Eq.~(\ref{shift}). On the other hand, Fig.~\ref{fig1}(b) is equal to the previous data with use of Eq.~(\ref{non}). The most important contrasts between them are the line shape and the temperature range of critical fluctuation, although both of them become efficient criteria for the ferromagnetic phase transition. The difference between $\rho_{f}$ and $\rho_{n}$ by $B$ clearly appears.

At first, we find that the quantity $S-(c/3)\ln L$ in Fig.~\ref{fig1}(a) takes a small positive constant value at $T_{c}$ and the constant seems to be independent on the system size $L$. Thus, we expect the presence of Eq.~(\ref{CFT}) at the critical point. Strikingly, the behavior of $S-(c/3)\ln L$ is similar to that of the Binder parameter, since the data intersect with each other near $T_{c}$. The intersection near $T_{c}$ is not seen in Fig.~\ref{fig1}(b).

Next we find that the change in $S-(c/3)\ln L$ near $T_{c}$ is abrapt in Fig.~\ref{fig1}(a). This feature is also different from that in Fig.~\ref{fig1}(b), where we have observed that $S$ only weakly changes far above $T_{c}$ and starts to decrease slightly above $T_{c}$, followed by a quite asymmetric tail below $T_{c}$. Because of these features, we could not exactly separate the critical behavior from the high-$T$ $\ln L$ feature originating from the random matrix theory in the previous work. In the present case, the high-$T$ and near-$T_{c}$ features are different and they can be identified independently.

As the third point, the data in Fig.~\ref{fig1}(a) are more symmetric at around $T_{c}$ than those in Fig.~\ref{fig1}(b). This may indicates the fundamental nature of the exact solution in which the divergence of the specific heat is symmetric near $T_{c}$. The critical fluctuation above $T_{c}$ may also suggest that the entropy detects the violation of the ferromagnetic order.

\subsection{Average Snapshot Entropy}

\begin{figure}[htbp]
\begin{center}
\includegraphics[width=7cm]{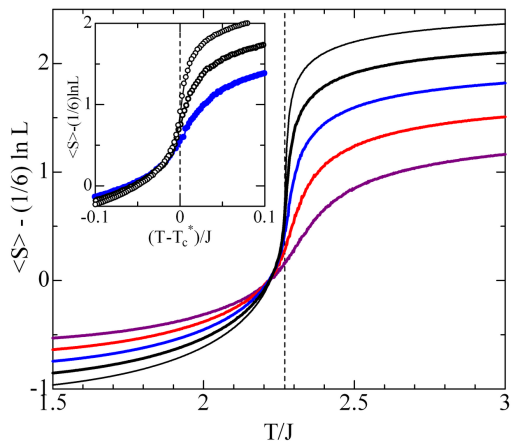}
\end{center}
\caption{(Color online) Temperature and system-size dependence on $\left<S\right>-(c/3)\ln L$ for $M(x,y)=(\sigma_{i}+1)/2$ with $i=(x,y)$: $L=32=2^{5}$ (purple), $L=64=2^{6}$ (red), $L=128=2^{7}$ (blue), $L=256=2^{8}$ (black), and $L=512=2^{9}$ (fine black line). A dashed line is a guide to $T_{c}$. The inset enlarges the data near $T_{c}$. Here we replot the data as a function of $T-T_{c}^{\ast}(L)$, where $T_{c}^{\ast}(L)$ is determined by Eq.~(\ref{peak}).}
\label{fig2}
\end{figure}

To look at the Binder-parameter-like feature more precisely, we take the sample average
\begin{eqnarray}
\left<S\right>=\frac{1}{N_{av}}\sum_{l=1}^{N_{av}}S(M_{l}),
\end{eqnarray}
where $S(M_{l})$ is equal to Eq.~(\ref{entropy}) for the single snapshot $M_{l}$. The index $l$ identifies the $l$th snapshot. We typically take $N_{av}=10^{3}\sim 10^{4}$ after MC steps at each $T$. We plot $\left<S\right>-(c/3)\ln L$ in Fig.~\ref{fig2}. We confirm that the intersection seems to occur at around $2.20J<T<2.25J$, slightly below $T_{c}$. We believe that this intersection is really due to the critical nature of the model, but at the same time the finite-size correction is necessary for the exact intersection at $T_{c}$. Usually, we may use a standard scaling plot based on $(T-T_{c})L^{1/\nu}$ and $L^{h}$ with the critical expoment $\nu$ and the scale dimension $h$. However, we should be careful for the fact that the expected $L$-dependence on $\left<S\right>$ is not power-law type. Thus, we need to introduce new scaling plot.

We replot the data as a function of $T-T_{c}^{\ast}(L)$ in the inset of Fig.~\ref{fig2}, where $T_{c}^{\ast}(L)$ is the peak position of the specific heat for finite size $L$. In the cluster algorithm, we define a temperature grid as $\Delta T=0.001\times T_{c}$, and we find $T_{c}^{\ast}(L=64)=2.2828$, $T_{c}^{\ast}(L=128)=2.2760$, and $T_{c}^{\ast}(L=256)=2.2714$ (The error is of order $\pm 0.002$, and the last digit is meaningless). These data agree well with the scaling for the peak position of the specific heat
\begin{eqnarray}
T_{c}^{\ast}(L)\sim T_{c}+\frac{D}{L^{1/\nu}}, \label{peak}
\end{eqnarray}
with a constant $D=0.87$ and the critical exponent $\nu=1$ for the Ising model. By this finite-size correction, the intersection point shifts toward $T=T_{c}^{\ast}(L)$. Thus, the numerical data suggest the presence of the asymptotic scaling given by
\begin{eqnarray}
\left<S\right>\sim\frac{c}{3}\ln L+\alpha, \label{sav}
\end{eqnarray}
with a positive constant $\alpha\sim 0.5$. The reason for the correction $T_{c}^{\ast}(L)$ is quite simple: the most fractal-like spin configuration appears at $T_{c}^{\ast}(L)$ for finite $L$, not at real $T_{c}$ in the thermodynamic limit.

\subsection{The Largest SVD Spectrum}

\begin{figure}[htbp]
\begin{center}
\includegraphics[width=7cm]{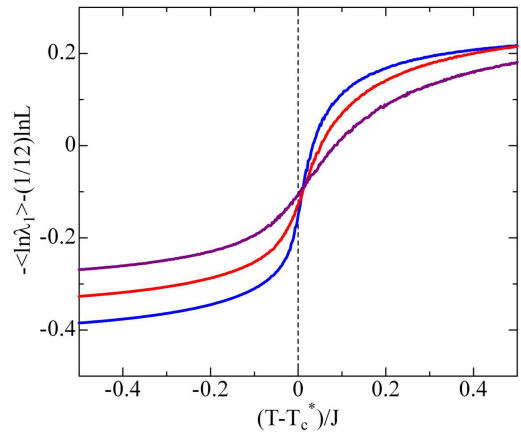}
\end{center}
\caption{(Color online) Temperature and system-size dependence on $-\left<\ln\lambda_{1}\right>-(c/6)\ln L$ for $M(x,y)=(\sigma_{i}+1)/2$ with $i=(x,y)$: $L=32=2^{5}$ (purple), $L=64=2^{6}$ (red), and $L=128=2^{7}$ (blue). We take $10^{5}$ MC steps and $N_{av}=10^{4}$.}
\label{fig3}
\end{figure}

Let us next examine whether Eq.~(\ref{spectrum}) is satisfied on the classical side. We calculate the $T$ dependence of $-\left<\ln\lambda_{1}\right>-(c/6)\ln L$. The result is shown in Fig.~\ref{fig3}. Here we take the sample average after taking the logarithm of $\lambda_{1}$. This may generate large sample deviation, since the logarithm for $\lambda_{1}<1$ changes quite rapidly. On the other hand, the obtained numerical data are not so seriously fructuating. Therefore, we can say that the data quality is retained. We find that the overall feature is similar to $\left<S\right>$, and we thus think that there is certain correspondence between the largest SVD spectrum and the largest entanglement spectrum. Near $T=T_{c}^{\ast}(L)$ (the error is the order of $0.01J$), we actually find that all the data intersect with each other, indicating the presence of the scaling given by
\begin{eqnarray}
-\left<\ln\lambda_{1}\right>\sim\frac{c}{6}\ln L+\alpha^{\prime}. \label{lav}
\end{eqnarray}
The additional negative constant $\alpha^{\prime}$ is considered to originate from the ambiguity of the background matrix. We observe $\alpha^{\prime}\sim -0.1$. The result for the SVD spectrum is in also nontrivial agreement with the entanglement spectrum of 1D quantum physics.

\begin{figure}[htbp]
\begin{center}
\includegraphics[width=7cm]{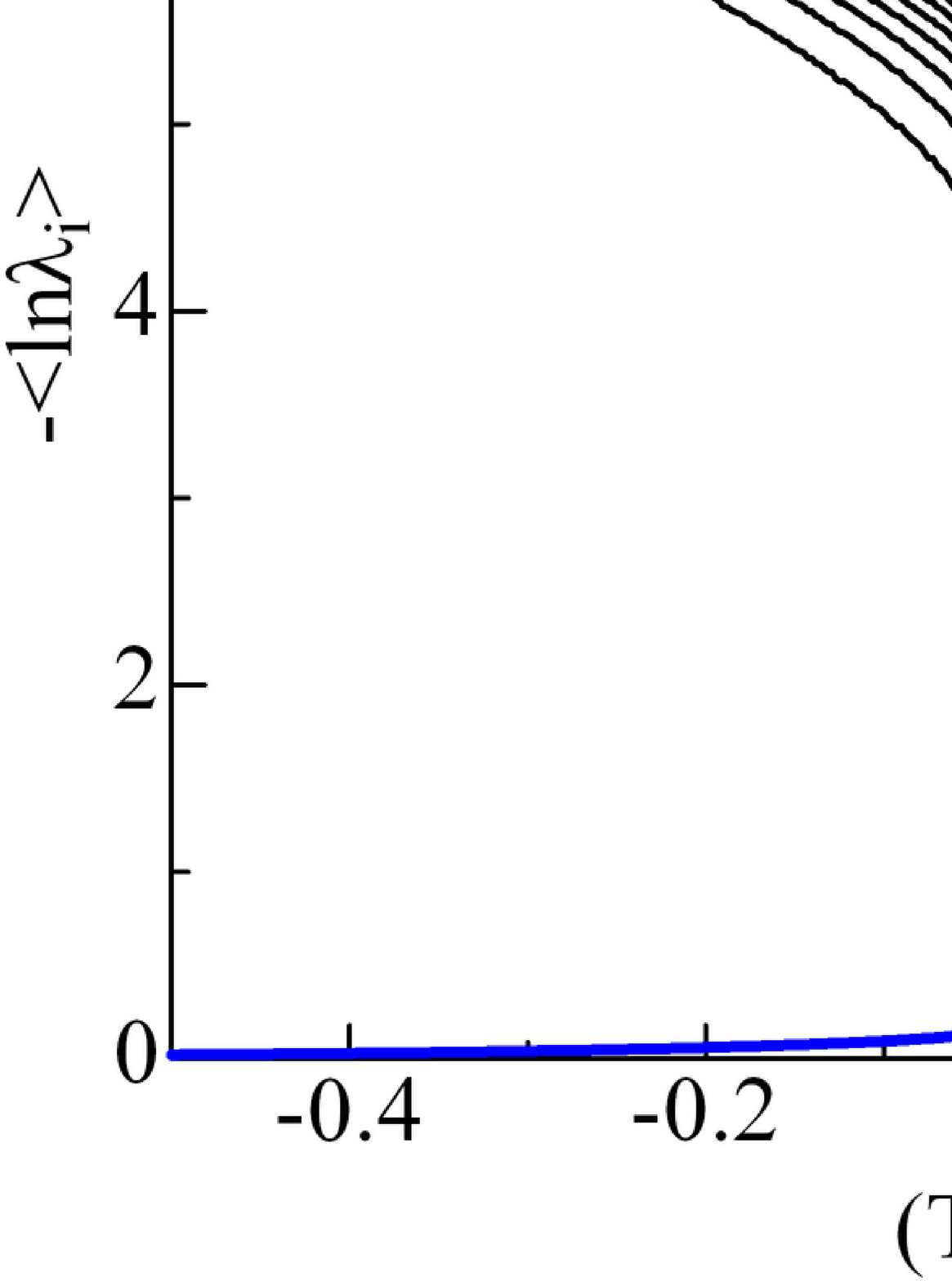}
\includegraphics[width=7cm]{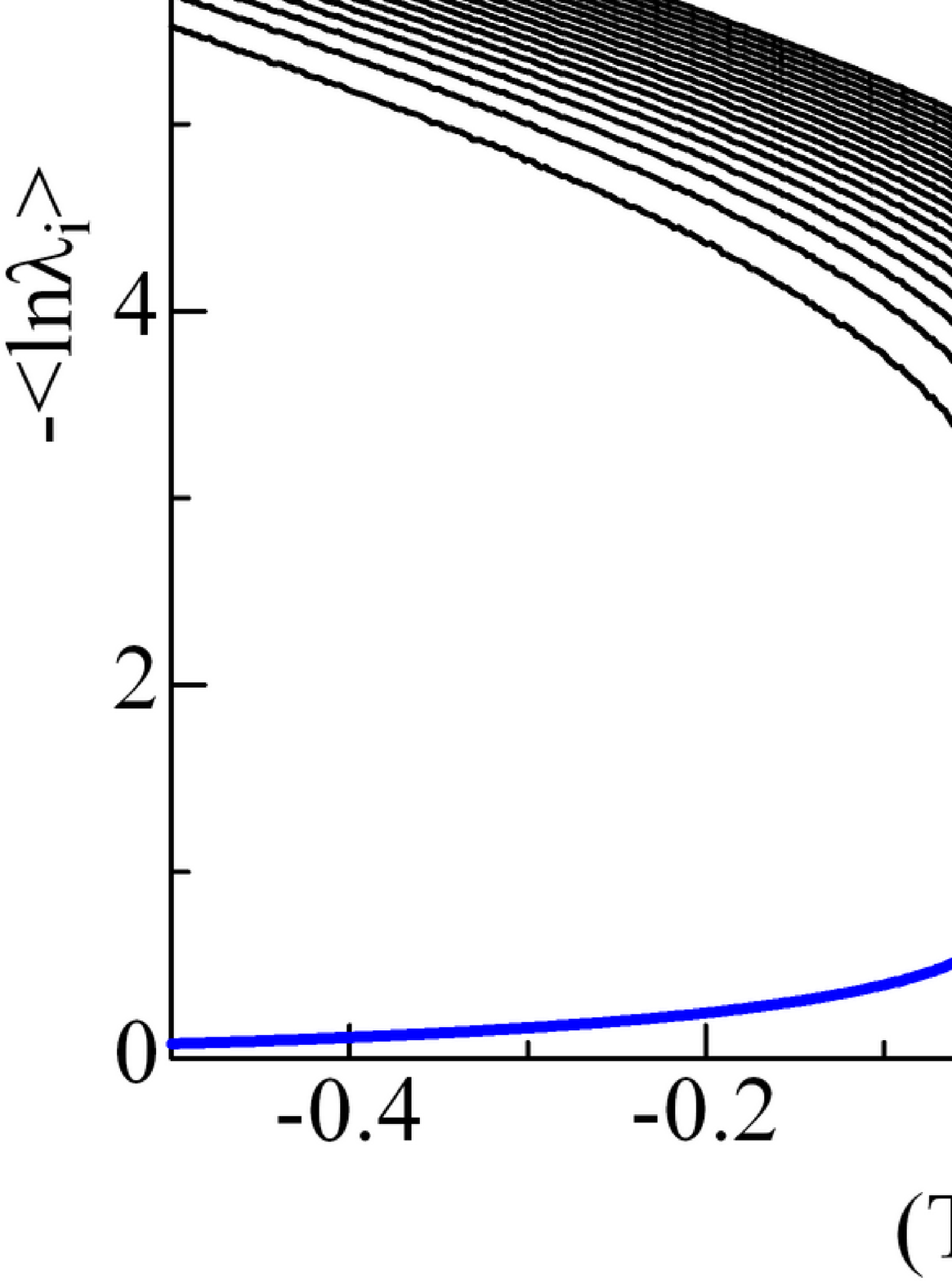}
\end{center}
\caption{(Color online) Temperature dependence on the SVD spectra for $L=128=2^{7}$. A blue line represents the largest spectrum $-\left<\ln\lambda_{1}\right>$. The other data above the gap are $-\left<\ln\lambda_{i}\right>$ for $i=2,3,...,20$ from bottom to top. (a) $M(x,y)=(\sigma_{i}+1)/2$, and (b) $M(x,y)=\sigma_{i}$.}
\label{fig4}
\end{figure}

To show peculiarity of the largest SVD spectrum, we plot the average spectra $-\left<\ln\lambda_{i}\right>$ ($\lambda_{1}>\lambda_{2}>\cdots>\lambda_{20}$) for $L=128$ in Fig.~\ref{fig4}(a). We observe that there is a large spectral gap between $-\left<\ln\lambda_{1}\right>$ and others (a band above the gap) in the ferromagnetic phase. We also observe that the gap tends to decrease toward $T_{c}$, and takes the minimum value. In the paramagnetic phase, the gap size shows weak temperature dependence. We have numerically confirmed that the decrease in the gap size at $T_{c}$ becomes remarkable with increasing $L$.

As we have already mentioned in Sec.II, $\lambda_{1}$ increases by the factorization, and then the average spectrum $-\left<\ln\lambda_{1}\right>$ decreases. As a result, this gap structure appears. Figure~\ref{fig4}(b) is the SVD spectrum before the factorization. Therein, the lift of the degeneracy between $-\left<\ln\lambda_{1}\right>$ and the continuous band is not complete above $T_{c}$. The mixture among these spectra blurs the essential properties of $-\left<\ln\lambda_{1}\right>$, when we calculate the snapshot entropy. The gap protects the critical behavior against the temperature fructuation. Therefore, the factorization plays crucial roles on the formation of the lowest spectrum separated from the continuous band.

\subsection{Advanced Scaling with Finite-$\beta$ on the Quantum Side and its Relation to $\beta_{c}$ of the Classical Ising Model}

Now it is possible to say that the snapshot entropy can detect the entanglement entropy of a 1D quantum system. We have already taken finite-$L$ correction on the classical side. However, we do not still take care about possible corrections to the formula $\left<S\right>\sim(c/3)\ln L$ itself. This is important, since the core factor of our scaling is whether the snapshot entropy can detect the proper entanglement entropy formula on the quantum side.

According to Refs.~\cite{Calabrese,Azeyanagi}, at finite inverse temperature $\beta$ (normalized by $J$), the entropy formula is given by
\begin{eqnarray}
S = \frac{c}{3}\ln\left(\frac{\beta}{\pi}\sinh\left(\frac{\pi L}{\beta}\right)\right) + c_{1}^{\prime}, \label{finiteS}
\end{eqnarray}
where $c_{1}^{\prime}$ is a nonuniversal constant. This formula becomes equivalent to Eq.~(\ref{CFT}) in the large $\beta$ limit, $\lim_{\beta\rightarrow\infty}S\sim(c/3)\ln L$. Thus, Eq.~(\ref{finiteS}) is generalization of Eq.~(\ref{CFT}). When we enlarge the inset in Fig.~\ref{fig2}, we find that the intersection at $T_{c}$ is not perfect. The intersection point still deviates from the exact $T_{c}$ by roughly $0.01J$. This ambiguity would be corrected by the $\beta$ scaling.

\begin{figure}[htbp]
\begin{center}
\includegraphics[width=7cm]{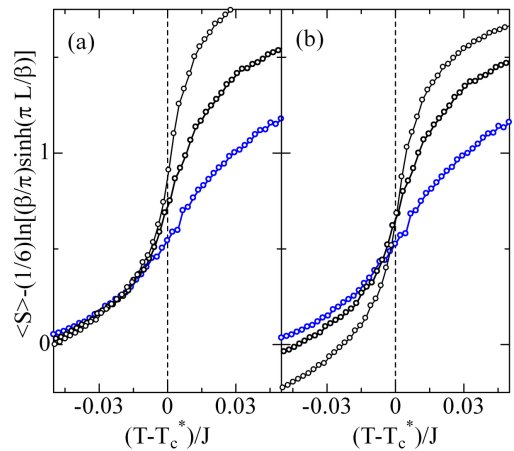}
\end{center}
\caption{(Color online) $\beta$-dependence of Eq.~(\ref{finiteS}) on the intersection. (a) $\beta\rightarrow\infty$ and (b) $\beta=500$. We take $L=128=2^{7}$ (blue), $L=256=2^{8}$ (black), and $L=512=2^{9}$ (fine black line and circles).}
\label{fig5}
\end{figure}

In Fig.~\ref{fig5}, we compare the result for $\beta\rightarrow\infty$ with that for $\beta=500$. Here we introduce the scaling function by based on Eq.~(\ref{finiteS}). This value, $\beta=500$, is just taken so that the data match with each other at around $T=T_{c}$. We observe that this finite $\beta$ value actually improves the intersection feature. This result suggests that more precise form than Eq.~(\ref{sav}) is given by
\begin{eqnarray}
\left<S\right>\sim\frac{c}{3}\ln\left(\frac{\beta}{\pi}\sinh\left(\frac{\pi L}{\beta}\right)\right)+\alpha.
\end{eqnarray}
Note that the data for $L=128$ are almost unchanged for $\beta=500$, and the data for $L=256$ and $L=512$ show almost parallel shift toward the $L=128$ data. Now, we are thinking about that the corresponding 1D quantum system is in the inverse temperature $\beta$. We would like to know the physical meaning of this $\beta$ value.

For that purpose, let us make the above point mathematically clear. We will see that the presence of $\beta$ is significant for the finite-size correction. Let us start with the 1D transverse-field quantum Ising model at the critical point. The Hamiltonian ($J=1$) is defined by
\begin{eqnarray}
H_{1D}=-\sum_{i}\sigma_{i}^{z}\sigma_{i+1}^{z}-\lambda\sum_{i}\sigma_{i}^{x}, \label{transverse}
\end{eqnarray}
where this model has also $c=1/2$, and the quantum phase transition occurs at $\lambda=1$ and $\beta\rightarrow\infty$. The partition function $Z$ is transformed into the following form by the Suzuki-Trotter decompostion
\begin{eqnarray}
Z&=&{\rm tr}\left(e^{-\beta H_{1D}}\right) \nonumber \\
&=&\lim_{M,\beta\to\infty}A^{M}\sum_{\{\sigma^{1}\}}\cdots\sum_{\{\sigma^{M}\}}\prod_{k=1}^{M}\exp\left\{\frac{\beta}{M}\sum_{i=1}^{L}\sigma_{i}^{k}\sigma_{i+1}^{k}\right\} \nonumber \\
&&\times\exp\left\{\frac{1}{2}\ln\left(\coth\frac{\beta}{M}\right)\sum_{i=1}^{L}\sigma_{i}^{k}\sigma_{i}^{k+1}\right\},
\end{eqnarray}
where $A=\sqrt{(1/2)\sinh(\beta/M)}$, $M$ is the Trotter number, and $\{\sigma^{k}\}$ represents a particular spin configuration with label $k$, $\{\sigma^{k}\}=\left(\sigma_{1}^{k},\sigma_{2}^{k},...,\sigma_{L}^{k}\right)$. The Trotter decomposition becomes exact in the large $M$ limit. Thus, it is quite important to take $\beta\rightarrow\infty$ and $M\rightarrow\infty$ simultaneously as well as keeping the ratio $\beta/M$ finite. Here we introduce $\beta_{c}$, the inverse critical temperature in the 2D classical isotropic Ising model
\begin{eqnarray}
\beta_{c}=\frac{1}{T_{c}}=\frac{1}{2}\ln\left(1+\sqrt{2}\right).
\end{eqnarray}
If the quantum criticality of Eq.~(\ref{transverse}) at $\beta\rightarrow\infty$ corresponds to phase transition in the classical side at $\beta_{c}$, we can rewrite this equation as
\begin{eqnarray}
Z=\lim_{M,\beta\to\infty}A^{M}\sum_{\{\sigma^{1}\}}\cdots\sum_{\{\sigma^{M}\}}e^{-\beta_{c}H_{2D}},
\end{eqnarray}
where the effective Hamiltonian is defined by
\begin{eqnarray}
H_{2D}=-J_{1}\sum_{k=1}^{M}\sum_{i=1}^{L}\sigma_{i}^{k}\sigma_{i+1}^{k}-J_{2}\sum_{k=1}^{M}\sum_{i=1}^{L}\sigma_{i}^{k}\sigma_{i}^{k+1},
\end{eqnarray}
with the effective interaction
\begin{eqnarray}
J_{1}=\frac{1}{\beta_{c}}\frac{\beta}{M} \; , \; J_{2}=\frac{1}{2\beta_{c}}\ln\left(\coth\frac{\beta}{M}\right).
\end{eqnarray}
In the present setup (2D square lattice), we should take $M=L$.

The above consideration is quite important, since in the holographic theory the correspondence is strongly restricted by the common symmetry between quantum and classical systems. Then, the critical point in a quantum theory should be mapped onto the corresponding critical point in a classical theory.

We should take $\beta\rightarrow\infty$ and $M\rightarrow\infty$ simultaneously so that their ratio, $x=\beta/M=\beta/L$, remains finite. In this special setup, we can obtain the isotropic 2D classical Ising model ($J_{1}=J_{2}$) by taking the condition
\begin{eqnarray}
x=\frac{1}{2}\ln\left(\coth x\right),
\end{eqnarray}
and the exact solution is found to be
\begin{eqnarray}
x=\frac{\beta}{L}=\beta_{c}=\frac{1}{2}\ln\left(1+\sqrt{2}\right)\sim 0.44. \label{x}
\end{eqnarray}
This solution leads to
\begin{eqnarray}
J_{1}=J_{2}=1.
\end{eqnarray}
When we substitute Eq.~(\ref{x}) into the scaling formula (\ref{finiteS}), we recover Eq.~(\ref{CFT}).

The constant ratio $x$ is necessary for $\beta\to\infty$ and for the correct Trotter decomposition ($M=L\to\infty$ for the square lattice). At the same time, we can think of approximately extracting the zero-temperature feature on the quantum side from finite-size data. Then we need to take a finite $\beta$ value with keeping $x$ constant. In the present case, we fix the $\beta$ value, and changes the finite size $L$. Thus $x$ also changes from $0.44$, and this change induces some correction to the approximate zero-temperature feature. Substituting $x=\beta/L$ into Eq.~(\ref{finiteS}), we can decompose $S$ into the following form
\begin{eqnarray}
S=\frac{c}{3}\ln L + \Delta S + c_{1}^{\prime},
\end{eqnarray}
and the possible correction $\Delta S$ is represented as
\begin{eqnarray}
\Delta S\left(x(L)\right)=\frac{c}{3}\ln\left(\frac{x}{\pi}\sinh\frac{\pi}{x}\right).
\end{eqnarray}
The data shift from Fig.~\ref{fig5}(a) to (b) is almost parallel along the perpendicular axis, and thus the shift is owing to the difference from zero-temperature entropy formula, $\Delta S\left(x(L)\right)$. We estimate $\Delta S\left(x(L=128)\right)=0.0176$, $\Delta S\left(x(L=256)\right)=0.0665$, and $\Delta S\left((L=512)\right)=0.2256$, and they are actually consistent with numerical data.

Before closing this section, we briefly mention the meaning of $c$ in our scaling. We have regarded the entanglement entropy formula in 1D as a relevant scaling equation of the snapshot entropy, and this identification was successfully observed by numerical simulation. Then, we have used $c=1/2$ of the 2D classical Ising model. In the present case, $c=1/2$ for both classical and quantum systems, and there is no ambiguity. However, in Sec.VI, we will examine the snapshot entropy of the three-states Potts model, where we can not find the corresponding quantum 1D system. Then, we necessarily use the central charge of the Potts model $c=4/5$ by considering that the criticality should be conserved after the quantum-classical correspondence. This problem may be similar to that in the string theory. Therein, the charge can be defined on the classical side, and is called as the Brown-Henneaux central charge~\cite{BH}, in which the charge is replaced by the curvature radius devided by the Newton constant on the classical side. Although our scaling formula comes from the quantum side, we think that it would be enough to take account of the central charge on the classical side. We will again mention this point in Sec.~VI.

\section{Tensor Product of Factorized Matrix and Cluster Distribution}

\subsection{Factorization and Cluster Distribution}

Let us consider the appearance of the central charge $c$ in more physical standpoint. We examine that the factorization changes the distribution of the number of the ferromagnetic cluster sizes in the snapshot. This is crucial for determining the proper spin-spin correlator form. To see this, we introduce an effective model defined by the following unit cell matrix
\begin{eqnarray}
H=\left(
\begin{array}{cccc}
1&1&0&1 \\
1&1&0&0
\end{array}
\right), \label{M01}
\end{eqnarray}
where the left half block ($4$ spins) represents a large spin cluster, while the right upper site reprensets a small cluster ($1$ spin). As already mentioned in the introduction, the fractal-like spin configuration can be made by the multiple tensor product of the unit cell $H$. Taking the tensor product between two copies of $H$, we obtain
\begin{eqnarray}
H\otimes H=\left(
\begin{array}{cccccccccccccccc}
1&1&0&1&1&1&0&1&0&0&0&0&1&1&0&1 \\
1&1&0&0&1&1&0&0&0&0&0&0&1&1&0&0 \\
1&1&0&1&1&1&0&1&0&0&0&0&0&0&0&0 \\
1&1&0&0&1&1&0&0&0&0&0&0&0&0&0&0
\end{array}
\right). \nonumber \\
\end{eqnarray}
To clearly look at cluster destribution, we omit zero components from $H\otimes H$ (it is also possible to omit $1$). Then result is following:
\begin{eqnarray}
H\otimes H\rightarrow\left(
\begin{array}{cccccccccccccccc}
1&1&&1&1&1&&1&&&&&1&1&&1 \\
1&1&&&1&1&&&&&&&1&1&& \\
1&1&&1&1&1&&1&&&&&&&& \\
1&1&&&1&1&&&&&&&&&&
\end{array}
\right).
\end{eqnarray}
Then, we find certain distribution of various-size clusters. Here, we define a set of up spins (labeled by $1$) connected along bond direction as a cluster. The largest cluster size is $10$. We observe a rough tendency that the number of the clusters increases with decreasing the cluster size. This tendency matches with the spin configuration realized by the MC simulation at $T_{c}$.

On the other hand, if we consider the non-factorized case, the situation changes. Let us next consider the unit cel matrix
\begin{eqnarray}
H=\left(
\begin{array}{cccc}
1&1&-1&1 \\
1&1&-1&-1
\end{array}
\right), \label{Mpm}
\end{eqnarray}
and calculate the tensor product. Then we obtain
\begin{eqnarray}
H\otimes H\rightarrow\left(
\begin{array}{cccccccccccccccc}
1&1&&1&1&1&&1&&&1&&1&1&&1 \\
1&1&&&1&1&&&&&1&1&1&1&& \\
1&1&&1&1&1&&1&&&1&&&&1& \\
1&1&&&1&1&&&&&1&1&&&1&1
\end{array}
\right),
\end{eqnarray}
where we have omitted $-1$. We clearly see that there are many large spin clusters and the distribution changes.

\subsection{Cluster Distribution and Central Charge}

Let us discuss a close relation between the cluster distribution and the coefficient of the logarithmic term in the snapshot entropy formula. According to Ref.~\cite{Matsueda2}, the snapshot entropy is given by
\begin{eqnarray}
S=N\sum_{j=1}^{r}\left(-\gamma_{j}\ln\gamma_{j}\right)=\frac{1}{3}C\ln L,
\end{eqnarray}
where
$r={\rm rank}\left(HH^{\dagger}\right)=2$ (we denote this as $j=\pm$), and $\gamma_{j}$ are the normalized eigenvalues of $HH^{\dagger}$. This equation tells us that there exist unit cell matrices with a particular set of $\gamma_{j}$ values that leads to a proper central charge.

We compare the $C$ value for Eq.~(\ref{M01}) with that for Eq.~(\ref{Mpm}). We find $\gamma_{\pm}=(1/2)\pm(\sqrt{17}/10)$ for Eq.~(\ref{M01}), and $\gamma_{\pm}=(1/2)\pm(1/4)$ for Eq.~(\ref{Mpm}). Now the system after $N$ tensor products is a rectangular shape $2^{N}\times 4^{N}$, and thus we identify a typical linear size to be $L^{2}=2^{3N}$ and $N=(2/3)\log_{2}L$. We then obtain $C=0.857420$ for Eq.~(\ref{M01}) and $C=1.622556$ for Eq.~(\ref{Mpm}). In comparison with the real central charge $c=1/2$, they are much larger. However, these trends may suggest that a better cluster configuration tends to provides us with a $C$ value close to the real $c$ value.

Let us go back to Eq.~(\ref{unit}). We told that the Sierpinski carpet seems to be the ideal representation of the fractal-like spin configuration at $T_{c}$ of the Ising model, in which white pixels are regarded as spin islands. Actually, this case provides us with $\gamma_{\pm}=(1/2)\pm(\sqrt{3}/4)$ and $C=0.671143$ which is much better than $C=0.857420$ and $1.622556$.

Alternatively, we can set up a unit cell matrix that represents a slightly-ordered but still critically fructuating spin configuration. For instance, we have
\begin{eqnarray}
H=\left(\begin{array}{cccc}1&1&1&1 \\ 1&0&1&1 \\ 1&1&1&1\\ 1&1&1&1\end{array}\right). \label{M01new}
\end{eqnarray}
The matrix $H^{2}$ has two non-zero eigenvalues $\gamma_{\pm}=(15\pm\sqrt{189})/2$. In this case, $L=4^{N}$ and we obtain $C=0.375337$ smaller than $c=1/2$. We think that the density of zero components providing the best configuration is between the density of zeros in Eq.~(\ref{M01}) and that in Eq.~(\ref{M01new}).

\section{Finite-$\chi$ Scaling}

Up to now we have confirmed that the factorization changes the entropy scaling very much. Thus, we are also interested in whether the coarse-grained snapshot entropy formula is also deformed or not. The coarse-grained entropy is defined by
\begin{eqnarray}
S_{\chi}=-\sum_{n=1}^{\chi<L}\lambda_{n}\ln\lambda_{n}.
\end{eqnarray}
It has been pointed out recently that the coarse-grained entropy obeys anomalous scaling that is somewhat different from the finite-entanglement scaling in quantum 1D systems near criticality, because the snapshot density matrix roughly corresponds to the two-point spin correlation function rather than a state itself~\cite{Imura,Tagliacozzo,Pollmann}. In this viewpoint, holographic transformation only appears on the full snapshot entropy. The finite-entanglement scaling~\cite{Tagliacozzo,Pollmann} is given by
\begin{eqnarray}
S_{\chi}=\frac{1}{\sqrt{12/c}+1}\ln\chi, \label{Pollmann}
\end{eqnarray}
and the recently proposed scaling is given by
\begin{eqnarray}
S_{\chi}=b\chi^{\eta}\ln\left(\frac{\chi}{a}\right), \label{Imura}
\end{eqnarray}
where $\eta$ is the anomalous dimension, and in the Ising case we have $\eta=1/4$~\cite{Imura}. This scaling is derived from algebraic decay of the singular value spectrum as
\begin{eqnarray}
\lambda_{n}=\frac{A}{n^{1-\eta}},
\end{eqnarray}
where $A$ is a constant fitting parameter~\cite{Matsueda3}.

\begin{figure}[htbp]
\begin{center}
\includegraphics[width=7cm]{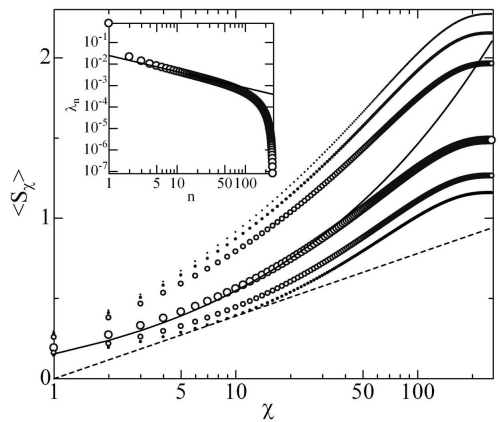}
\end{center}
\caption{(Color online) $\left<S_{\chi}\right>$ for $L=256$ at $T=2.30J$, $2.29J$, $2.28J$, $2.27J$, $2.26J$, and $2.25J$ (from top to bottom). Solid and dashed lines denote Eq.~(\ref{Imura}) with $a=0.1$ and $b=0.067$ and Eq.~(\ref{Pollmann}), respectively. The inset represents the snapshot spectrum at $T=2.27J$. A guide line denotes $\lambda_{n}=A/n^{1-\eta}$ with $A=0.025$.}
\label{fig6}
\end{figure}

Figure~\ref{fig6} shows $S_{\chi}$ for $L=256$ near $T_{c}$. We have averaged $S_{\chi}$ over $N_{av}=10^{5}$ samples, and denote the results as
\begin{eqnarray}
\left<S_{\chi}\right>=\frac{1}{N_{av}}\sum_{l=1}^{N_{av}}S_{\chi}(M_{l}).
\end{eqnarray}
We find that the numerical data fit with Eq.~(\ref{Imura}) for wide-$\chi$ range. We also show the snapshot spectrum in the inset of Fig.~\ref{fig3}. The result is consistent with the recent works~\cite{Imura,Matsueda3}. These results show that the anomalous scaling with $\eta$ is robust against the factorization. As we have mentioned in Sec.II, only the largest eigenvalue $\lambda_{1}$ increases by the factorization. Thus if we forget the normalization, the finite-$\chi$ dependence is not affected. Actually, the inset clearly shows the presence of the algebraic decay of $\lambda_{n}$ ($1<n\ltsim 100$) that is the origin of the anomalous scaling Eq.~(\ref{Imura}), and then $\lambda_{1}$ largely deviates from this scaling.

\section{Snapshot Entropy for Three-States Potts Model}

Finally, we discuss about the snapshot entropy for the three-states Potts model. The Hamiltonian is given by
\begin{eqnarray}
H=-J\sum_{<i,j>}\delta\left(\sigma_{i},\sigma_{j}\right),
\end{eqnarray}
where $\sigma_{i}$ is usually taken to be $-1$, $0$, and $1$. The central charge of the model is $c=4/5$, and thus our holographic conjecture can be well supported by wider universality classes. A proper description of $M_{f}$ may be given by
\begin{eqnarray}
M_{f}=M_{n}+B,
\end{eqnarray}
and $B$ in Eq.~(\ref{b}) is used here. We obtain snapshots by the MC simulation, and apply SVD to the snapshots to calculate the entropy.

\begin{figure}[htbp]
\begin{center}
\includegraphics[width=7cm]{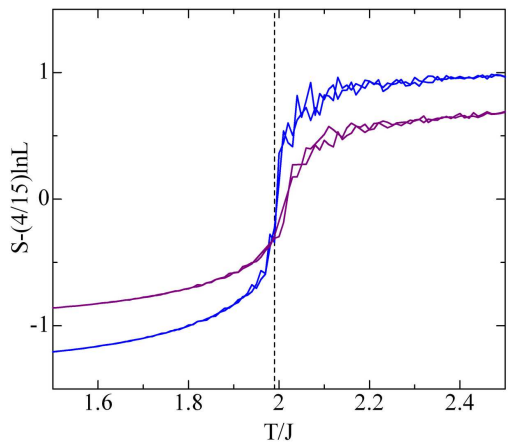}
\end{center}
\caption{(Color online) $\left<S\right>-(c/3)\ln L$ with use of the factorized form $M_{f}=M_{n}+B$ as a function of $T$ for the three-states Potts model: $L=32=2^{5}$ (purple) and $L=128=2^{7}$ (blue). Two lines with the same color represent different runs. A dashed line is a guide to $T_{c}$.}
\label{fig7}
\end{figure}

In Fig.~\ref{fig7}, we plot $\left<S\right>-(c/3)\ln L$ for $L=32$ and $L=128$ as a function of $T$. We observe the intersection at around $T_{c}=2J/\ln\left(1+\sqrt{3}\right)=1.98994J$. The result suggests that the scaling formula given by Eq.~(\ref{sav}) is also satisfied for $c=4/5$, but it is interesting that $\alpha$ is now negative. The result also suggests that the central charge in the scaling formula can be defined by the central charge on the classical side, although in the Potts model case there is no information about the corresponding 1D quantum system. Here, we did not consider the difference between $T_{c}$ and $T_{c}^{\ast}(L)$. This is because the transition is quite sharp even for relatively small sizes of order $L=128$. According to Eq.~(\ref{peak}), we think that $D$ must be quite small, because $\nu=5/6$ for the three-states Potts model and there is the $L$-dependence on $T_{c}^{\ast}(L)$. By combining $\alpha<0$ with $\beta\rightarrow\infty$ together, we imagine that the corresponding 1D system might be the topological ground state of a quantum model with $c=4/5$. The confirmation of this conjecture is an interesting future work.

\section{Summary}

In summary, we have examined the role of the factorized snapshot matrix on the entropy scaling in the 2D classical Ising model. The result is quite surprizing and seems to contain enough information of the holographically connecting 1D quantum system, including finite-temperature efffects, although we just replace $-1$ into $0$ in the snapshot data. This result has been also confirmed for different universality classes by calculating the snapshot entropy of the three-states Potts model. The point behind this factorization is how to correctly treat the largest singular value of the snapshot matrix. This point has been also confirmed numerically by calculating the largest singular spectrum. Furthermore, the distribution of the cluster sizes in the snapshot matrix is related to the coefficient of the logarithmic scaling formula. This means that the coefficient of the logarithmic term is actually related to the critical properties of the system. This critical property agrees well with the entanglement entropy formula in 1D critical systems.

The snapshot may seem to be different from physical quantities. However, it is almost obvious that the average entropy should have enough information of the partition function. The present result shows that the entropy detects the total amount of information of the system, not the quality of the snapshot, and thus the holographic connection between completely different systems naturally emerges from the present entropic calculations.

H.M. acknowledges Tsuyoshi Okubo and Ching Hua Lee for enlightening discussions.

\end{document}